\documentclass[journal,onecolumn]{IEEEtran} %%% USE THIS FOR FINAL PAPER

\makeatletter
\let\NAT@parse\undefined
\makeatother
\usepackage[square,comma,sort&compress]{natbib} 
\usepackage{amsmath}
\usepackage{amssymb}
\usepackage{epsfig}
\usepackage{subfigure}
\usepackage{amsfonts}
\usepackage{amssymb}
\usepackage{latexsym}
\usepackage{diagrams}%
\usepackage{color}		% Need the color package for the figure

\newarrow{ADots}{<}{.}{.}{.}{>} % to use in diagram, us rADots, uADots, etc...

\newtheorem{theorem}{Theorem}

\newtheorem{definition}{Definition}

\newcommand{\ip}[2]{\langle#1,#2\rangle}

\newcommand{\lc}{{\cal L}}
\newcommand{\M}{{\mathfrak M}}

\newcommand{\Langle}{\left\langle}
\newcommand{\Rangle}{\right\rangle}

\newcommand{\Ran}{\text{Ran}\,}
\newcommand{\sinc}{\text{sinc}}

\newcommand{\Lop}{\mathcal{L}}
\newcommand{\sut}{\text{ s.t.}}

\newcommand{\Nop}{\mathcal{N}}
\newcommand{\Wop}{\mathcal{W}}

\newcommand{\F}{\mathcal{F}}

\newcommand{\Z}{\mathbb{Z}}
\newcommand{\R}{\mathbb{R}}

\newcommand{\dif}{\mathrm{d}}

\newcommand{\dt}{\dif t}

\newcommand{\da}{\dif a}
\newcommand{\db}{\dif b}

\newcommand{\dtau}{\dif \tau}

\newcommand{\dtheta}{\dif \theta}
\newcommand{\dnu}{\dif \nu}
\newcommand{\domega}{\dif \omega}

\newcommand{\doubint}{\int\!\!\!\!\int}
\newcommand{\tripint}{\int\!\!\!\!\int\!\!\!\!\int}

\newcommand{\defeq}{:=}
\newcommand{\qeq}{&=&}

\begin{document}

\title{Canonical time-frequency, time-scale, and frequency-scale
representations of time-varying channels\authorrefmark{1}\footnote{\authorrefmark{1}This research was supported in part by the Office of Naval Research under Grant N00014-03-1-0102.}}

\author{Scott T. Rickard\authorrefmark{2}\footnote{\authorrefmark{2}School of Electrical, Electronic and Mechanical Engineering, University College Dublin, Ireland, Email: scott.rickard@ucd.ie},
Radu V. Balan\authorrefmark{3}\footnote{\authorrefmark{3}Siemens Corporate Research, Princeton, New Jersey, USA.}, 
H. Vincent Poor\authorrefmark{4}\footnote{\authorrefmark{4}Department of
Electrical Engineering, Princeton University, Princeton, New Jersey,
USA.}, 
Sergio Verd\'{u}\authorrefmark{4}}

\maketitle

\begin{abstract}
Mobile communication channels are often modeled as linear time-varying
filters or, equivalently, as time-frequency integral operators with
finite support in time and frequency. Such a characterization
inherently assumes the signals are narrowband and may not be
appropriate for wideband signals. In this paper time-scale
characterizations are examined that are useful in wideband
time-varying channels, for which a time-scale integral operator is
physically justifiable. A review of these time-frequency
and time-scale characterizations is presented. Both the time-frequency and
time-scale integral operators have a two-dimensional discrete
characterization which motivates the design of time-frequency or
time-scale rake receivers. These receivers have taps for both time and
frequency (or time and scale) shifts of the transmitted signal. A
general theory of these characterizations which generates, as specific
cases, the discrete time-frequency and time-scale models is presented
here. The interpretation of these models, namely, that they can be
seen to arise from processing assumptions on the transmit and receive
waveforms is discussed. Out of this discussion a third model arises: a
frequency-scale continuous channel model with an associated discrete
frequency-scale characterization.
\end{abstract}


\IEEEpeerreviewmaketitle

\section{Introduction}
It is common to assume that a received communication signal is
composed of superpositions of different versions of the transmitted
signal.  These different versions arise from reflections of the signal
off of scatterers in the environment. In the {\em time-scale channel
model}, each reflection is a delayed and time scaled copy of the
transmitted signal. The delays arise from differing path lengths from
transmitter to scatterer to receiver. Relative motion of the
transmitter, scatterers, or receiver causes time
dilations/contractions of the transmitted waveform $x(t)$. Thus, each
reflection is of the form,
\begin{equation}
x_{a,b}(t) = \frac{1}{\sqrt{|a|}}x\left(\frac{t-b}{a}\right)
\end{equation}
and the received signal is a summation of the reflections
characterized by $\Lop(a,b)$, the {\em wideband spreading
function}\footnote{We will assume that all integrals are taken over
$(-\infty,\infty)$ unless otherwise specified.},
\begin{equation}\label{firstL}
y(t)=\doubint \Lop(a,b)\frac{1}{\sqrt{|a|}}x\left(\frac{t-b}{a}\right)\da\db.
\end{equation}

We call a time-scale channel a {\em wideband channel} when the
wideband spreading function has finite support. Due to the physical
limitations of signal propagation, it is reasonable to expect that
$\Lop(a,b)$ has finite support. The maximum possible rate of change in
path length, which is constrained by the speeds of the objects in the
environment, limits the support of $\Lop(a,b)$ to a narrow range
around the $a=1$ line. Causality and the propagation loss associated
with increasing path length effectively limit the support of
$\Lop(a,b)$ to a finite range in the $b$ direction. The support in the
$a$ direction causes a spreading in scale of the transmitted signal,
and the support in the $b$ direction causes a spreading in time of the
transmitted signal.  Thus, channels described by (\ref{firstL}) are
often referred to as {\em doubly spread} channels.

Many signals and signaling environments satisfy the {\em narrowband
condition}, an assumption under which the time dilations or
contractions are modeled as Doppler shifts.  Under this assumption,
each received reflection of the signal is assumed to be of the form,
\begin{equation}
x_{\tau,\theta}(t) = x(t-\tau)e^{j2\pi\theta t}\label{narrowecho}
\end{equation}
In the {\em narrowband channel model}, the received signal is a
superposition of time delayed and frequency shifted copies of the
input and the channel is characterized by the {\em narrowband
spreading function} $S(\theta,\tau)$,
\begin{equation}
y(t) = \doubint S(\theta,\tau)x(t-\tau)e^{j2\pi\theta t}\dtau\dtheta,\label{nc}
\end{equation}
where $S(\theta,\tau)$ typically has finite support in $\theta$ and
$\tau$ due to the physical limitations of the channel.  The span of
this spreading in time and frequency has proven to be a crucial
parameter in communication
systems~\cite{kailath62,bello63,kozek97transfer}.  Regardless of
whether the support constraint is satisfied or not, (\ref{nc}) is a
time-frequency description of a general time-varying linear system,
\begin{equation}\label{eq:1}
y(t) = \int h(t,\tau)x(t-\tau)\dtau.
\end{equation}
When $S(\theta,\tau)$ has no support constraint, the transmitted
waveform and environment need not satisfy the narrowband condition.

Kailath's pioneering work in his 1959 Master's thesis~\cite{kailath59}
and the concomitant development of the rake channel model provided a
mathematical framework for capturing the energy associated with
multiple transmission paths between transmitter and receiver using a
discretization of the channel model.  This work was furthered in
1963 by Bello, who proposed a discrete time-frequency characterization
of the time-varying channel~\cite{bello63}.  In~\cite{sayeed99joint},
Sayeed and Aazhang reinterpreted this characterization from a
diversity viewpoint, and used this canonical time-frequency channel
characterization which combines a discrete set of time delayed and
frequency shifted versions of the input signal,
\begin{equation}\label{intro-tfdiscrete}
y(t)  = \sum_{n=0}^{N}\sum_{k=-K}^{K}\hat{S}\left(\frac{k}{T},\frac{n}{W}\right)x\left(t-\frac{n}{W}\right)e^{j2\pi kt/T},
\end{equation}
where
\begin{equation}
\hat{S}(\theta,\tau)\defeq
\doubint S(\theta^\prime,\tau^\prime)\sinc\left(\left(\tau-\tau^\prime\right)W\right)\sinc\left(\left(\theta-\theta^\prime\right)T\right) e^{-j\pi(\theta-\theta^\prime)T}\dtheta'\dtau'\label{intro-smoothS}
\end{equation}
to define a delay-Doppler RAKE receiver, a two-dimensional extension
of the classic rake receiver. The delay-Doppler rake takes advantage
of the inherent added channel diversity associated with time-varying
narrowband channels~\cite{sayeed99joint}.  While the narrowband
assumption is satisfied in many wireless communication signal
environments, many wireless systems are wideband due to the higher
data rates and multiaccess techniques~\cite{matz02}. Thus we may
expect, in light of differences in the narrowband and wideband models,
some advantages to the development of a canonical time-scale channel
characterization in wideband communication scenarios. Motivated by
this,~\cite{rickardphd,balan04} used the channel in
(\ref{firstL}) to derive a time-scale canonical channel model
\begin{equation}\label{eq:6}
y(t) = \sum_{m,n} \frac{c_{m,n}}{a_0^{m/2}}x\left(\frac{t-nb_0a_0^m}{a_0^m}\right).
\end{equation}
where $a_0,b_0$ are related to channel and signal characteristics, and
\begin{equation}\label{eq:7}
c_{m,n} = \doubint\lc(a,b) \sinc\left(m - \frac{\ln\,a}{\ln\,a_0}\right)\sinc\left(n -\frac{b}{ab_0}\right) \da\db
\end{equation}
An identical formula has been derived independently
in~\cite{jiang03,jiang05}.  There is a difference, however, in the
physical meaning of the decomposition in (\ref{eq:6})
between~\cite{rickardphd,balan04} and~\cite{jiang03,jiang05}. We will
discuss this difference in Section~\ref{interpsec} where we will also
present our point of view on canonical channel models. For us, a
canonical model will refer to a time-varying linear system applied to
a particular class of transmit signals whose output is measured
through a particular observation procedure.  For example, the
time-frequency canonical model derived in~\cite{sayeed99joint} is
based on bandlimited transmit signals observed at the receiver over a
finite observation horizon (i.e., a time-limited receiver). As
we discuss below, the time-scale canonical model can be derived from
bandlimited transmit signals being observed at a scale-limited
receiver. Furthermore, the new third canonical frequency-scale channel
model introduced in this paper can be derived from scale-limited
transmit signals being received at a time-limited receiver. We
elaborate on this point in Section~\ref{interpsec}.

Based on the above interpretation, in Section~\ref{sfm} we introduce a
frequency-scale time-varying channel model of the form:
\begin{equation}
\label{eq:8}
y(t) = \int_{-\infty}^{\infty} \int_0^{\infty} \hat{\rho}(\omega,a)
e^{j2\pi\omega t}\frac{1}{\sqrt{a}}x\left(\frac{t}{a}\right)\da\domega
\end{equation}
which is equivalent to (\ref{eq:1}) for positive time supported input
signals and positive time horizon receivers, as we show in Appendix
\ref{sec:equiv}.  The canonical channel model derived from (\ref{eq:8}) is
\begin{equation}\label{eq:9}
y(t) = \sum_{m,n}c_{m,n}e^{j2\pi mt/(T_2-T_1)}1_{[T_1,T_2]}(t)a_0^{n/2}
x(a_0^n t)
\end{equation}
where
\begin{equation}
1_{[T_1,T_2]}(t) = \left\{
\begin{array}{r@{\quad:\quad}l} 1 & T_1 \leq t \leq T_2 \\
0 & \text{otherwise}\end{array}\right.
\end{equation}
and $c_{m,n}$ are coefficients which depend on the span of the observation
time horizon ($T_2-T_1$), the scale domain bandwidth, and frequency-scale
spreading function (see Equation (\ref{eq:rmn})).

Each of the three doubly spread canonical channel models discussed
above motivates the development of a different two-dimensional rake
receiver. A delay-dilation rake receiver based on the canonical
time-scale channel characterization~\cite{jiang03,margetts04,jiang05}
leverages the diversity in wideband signaling environments in the same
way that the delay-Doppler rake leverages the diversity in narrowband
signaling environments~\cite{sayeed99joint}.  Such a channel model and
receiver may be particularly useful for ultra-wideband signaling due
to the extremely wide transmission signal bandwidth~\cite{win02,cassioli03}.

\subsection{Outline of paper}
In Section \ref{narrowc} we review background material on continuous
narrowband (time-frequency) and wideband (time-scale) channel
characterizations and examine simple one-path delay-Doppler and
one-path delay-dilation channels in the framework of these
representations. We derive and discuss the mapping between
time-frequency and time-scale kernel operators and note that there
exist time-frequency channels with no corresponding time-scale
channel.  In Section \ref{tfrake} we develop a general technique for
the generation of canonical channel models and demonstrate the
application of the technique to time-frequency and time-scale kernel
operators. In Section \ref{interpsec} we discuss the interpretation
and derivation of these canonical models from reasonable processing
assumptions on the transmit and receive waveforms. In Section
\ref{sfm} we propose a frequency-scale canonical channel
characterization based on the translation operators in frequency and
scale. We conclude and propose future work in Section \ref{sum}.

\section{Continuous Narrowband and Wideband Channel Characterizations}\label{narrowc}
In this section we review and discuss the time-frequency and
time-scale channel models and examine some simple channels to gain
some intuition concerning the characterizations. The time-frequency
description is a general time-varying linear system
characterization. However, in a slight abuse of nomenclature, we will
refer to all channel characterizations which can be related to the
channel described by $S(\theta,\tau)$ via Fourier transforms and phase
factors as narrowband channels. Specifically, in this section, we
discuss twelve such equivalent characterizations which were first explored
by Kailath~\cite{kailath59}, Zadeh~\cite{zadeh61}, and
Bello~\cite{bello63}. We call these ``narrowband'' characterizations
because when $S(\theta,\tau)$ has finite support, the characterization
is typically used only in narrowband systems and is not appropriate
for wideband signals. We will only discuss the support condition
constraint on $S(\theta,\tau)$ for the narrowband characterizations
when relevant, and consider the more general case where there is no
such constraint on the support of $S(\theta,\tau)$.  Similarly, we
will refer to channel characterizations based on the time-scale kernel
$\Lop(a,b)$ as wideband characterizations because they are typically
used in a wideband setting~\cite{sibul94}.

\subsection{Narrowband Characterizations}
In this section, we
develop a general technique for the generation of canonical channel
models and demonstrate the application of the technique to
time-frequency and time-scale kernel operators. 

The linear time-varying channel is characterized by the time-varying
impulse response $h(t,\tau)$ which denotes the response of the channel
at time $t$ to an impulse at time $t-\tau$. The channel input-output
relationship is thus,
\begin{equation}\label{tvlc}
y(t) = \int h(t,\tau)x(t-\tau)\dtau
\end{equation}
Such notation is used in, for
example,~\cite{bene99,biglieri98fading,proakis84,vantrees71,sayeed99joint}.

Another possible notation for the time-varying impulse response is
\begin{equation}
y(t) = \int k_0(t,\tau)x(\tau)\dtau.
\end{equation}
with the interpretation that $k_0(t,\tau)$ is the response of the
channel at time $t$ to an impulse at time $\tau$. This is the
formulation used in, for
example,~\cite{zadeh50,verdu98,matz2002LTV}. Bello~\cite{bello63}
calls $k_0(t,\tau)$ a {\em kernel system function} and notes the
obvious correspondence between the two representations,
$h(t,\tau)=k_0(t,t-\tau)$.  Bello~\cite{bello63} defines four
equivalent representations of the time-varying channel represented by
$k_0(t,\tau)$ that map the time or frequency representations of the
input into the time or frequency representations of the output. We
define these four kernel functions,
\begin{equation}\label{fourks}
\begin{aligned}
y(t) &= \int k_0(t,\tau)x(\tau)\dtau & \quad Y(\theta) & = \int k_1(\theta,\tau)x(\tau)\dtau\\
y(t) &= \int k_2(t,\nu)X(\nu)\dnu & \quad Y(\theta) &= \int k_3(\theta,\nu)X(\nu)\dnu
\end{aligned}
\end{equation}
The kernel system functions can be transformed into one another using
the Fourier transform. For example, the kernel function that maps the
input time domain to the output time domain ($k_0(t,\tau)$) and the
kernel function that maps the input time domain to the output
frequency domain ($k_1(\theta,\tau)$) are Fourier transforms of one
another with respect to the first argument. We can summarize the
relationships among the kernel system functions as follows,
\begin{equation}\label{krels}
\begin{diagram}
k_0(t,\tau)            & \rTo^{\F_{t\to\theta}} &  k_1(\theta,\tau)\\
\uTo^{\F_{\nu\to\tau}} &                        &  \uTo_{\F_{\nu\to\tau}}\\
k_2(t,\nu)             & \rTo_{\F_{t\to\theta}} &  k_3(\theta,\nu)
\end{diagram}
\end{equation}
That is,
\begin{equation}
\begin{aligned}
k_0(t,\tau) &= \int k_2(t,\nu)e^{-j2\pi \nu\tau}\dnu &\quad k_1(\theta,\tau) &= \int k_0(t,\tau)e^{-j2\pi t\theta}\dt\\
k_2(t,\nu) &= \int k_3(\theta,\nu)e^{j2\pi \theta t}\dtheta &\quad k_3(\theta,\nu) &= \int k_1(\theta,\tau)e^{j2\pi \tau\nu}\dtau
\end{aligned}
\end{equation}
The direction of the Fourier transform between $k_0$ and $k_2$ (and also between $k_1$ and $k_3$) is opposite to convention; We take the Fourier transform with respect to a ``frequency'' variable ($\nu$) and replace it with a ``time'' variable ($\tau$). This is necessary to be consistent with the kernel functions as defined in (\ref{fourks}).

Bello~\cite{bello63} provides the following useful interpretation of the kernel system functions,
\begin{itemize}
\item The response to input $\delta(t-t_0)$ is time function $k_0(t,t_0)$ with spectrum $k_1(\theta,t_0)$,
\item The response to input $e^{j2\pi\theta_0 t}$ is time function $k_2(t,\theta_0)$ with spectrum $k_3(\theta,\theta_0)$,
\end{itemize}
and also notes, by simple inspection of (\ref{fourks}), that $k_0$ and
$k_3$ are time-frequency duals of one another, as are $k_1$ and
$k_2$. 

Despite the simple input-output interpretations, the kernel system
functions often lack intuitive physical
interpretations~\cite{kailath59}. For this reason, Bello~\cite{bello63} and Kailath~\cite{kailath63}
examined eight other system function characterizing the linear
time-varying channel. These eight system functions are (\ref{tvlc}); its time-frequency dual,
\begin{equation}
Y(\theta) = \int G(\theta,\nu)X(\theta-\nu)\dnu;
\end{equation}
the three functions obtained by taking
the Fourier transform of $h(t,\tau)$ with respect to $t$, $\tau$, and
both $t$ and $\tau$; and 
the three functions obtained by taking
the inverse Fourier transform of $G(\theta,\nu)$ with respect to $\theta$, $\nu$, and
both $\theta$ and $\nu$. These eight functions are listed in Table~\ref{eight}.
\begin{table}
\begin{center}
\[
\begin{array}{|c|c|c|}\hline
h(t,\tau) & \mbox{\footnotesize input delay spread function} &  {\displaystyle y(t) = \int h(t,\tau)x(t-\tau)\dtau}\\\hline
S(\theta,\tau) & \mbox{\footnotesize delay-Doppler spreading function} & {\displaystyle S(\theta,\tau) =\int h(t,\tau) e^{-j2\pi t\theta}\dt}\\\hline
T(t,\nu) & \mbox{\footnotesize time-varying transfer function} & {\displaystyle T(t,\nu) =\int h(t,\tau) e^{-j2\pi \tau\nu}\dtau}\\\hline
H(\theta,\nu) & \mbox{\footnotesize output Doppler spread function} & {\displaystyle H(\theta,\nu) =\doubint h(t,\tau) e^{-j2\pi(t\theta+\tau\nu)}\dt\dtau}\\\hline
G(\theta,\nu) & \mbox{\footnotesize input Doppler spread function} & {\displaystyle Y(\theta)=\int G(\theta,\nu)X(\theta-\nu)\dnu}\\\hline
V(t,\nu) & \mbox{\footnotesize Doppler-delay spreading function} & {\displaystyle V(t,\nu) =\int G(\theta,\nu)e^{j2\pi\theta t}\dtheta}\\\hline
M(\theta,\tau) &\mbox{\footnotesize frequency dependent modulation function} & {\displaystyle M(\theta,\tau) =\int G(\theta,\nu)e^{j2\pi \nu\tau}\dnu}\\\hline
g(t,\tau) &\mbox{\footnotesize output delay spread function} & {\displaystyle g(t,\tau)=\doubint G(\theta,\nu) e^{j2\pi(\theta t+\nu\tau)}\dtheta\dnu}\\\hline
\end{array}
\]
\caption{Eight system functions characterizing the linear
time-varying channel, their function names from Bello~\cite{bello63}, and their associated input-output relationship or definition.}\label{eight}
\end{center}
\end{table}
In the current literature, $h(t,\tau)$ is usually referred to as the {\em time-varying impulse response}, 
(e.g.,~\cite{bene99,biglieri98fading,proakis84,vantrees71,sayeed99joint}) and the delay-Doppler spreading function, $S(\theta,\tau)$, is known simply as the {\em spreading function} (e.g.,~\cite{bene99,biglieri98fading,proakis84,vantrees71,sayeed99joint,verdu98}).
Unfortunately, $k_0(t,\tau)$ is also commonly referred to as the time-varying impulse response (e.g.,~\cite{zadeh50,verdu98}). We will refer to  $k_0(t,\tau)$  as the time-varying impulse response {\em kernel} to avoid confusion.

The relationships among the eight functions via duality and the Fourier transform are summarized in the following diagram. Duality is represented by a dotted line.
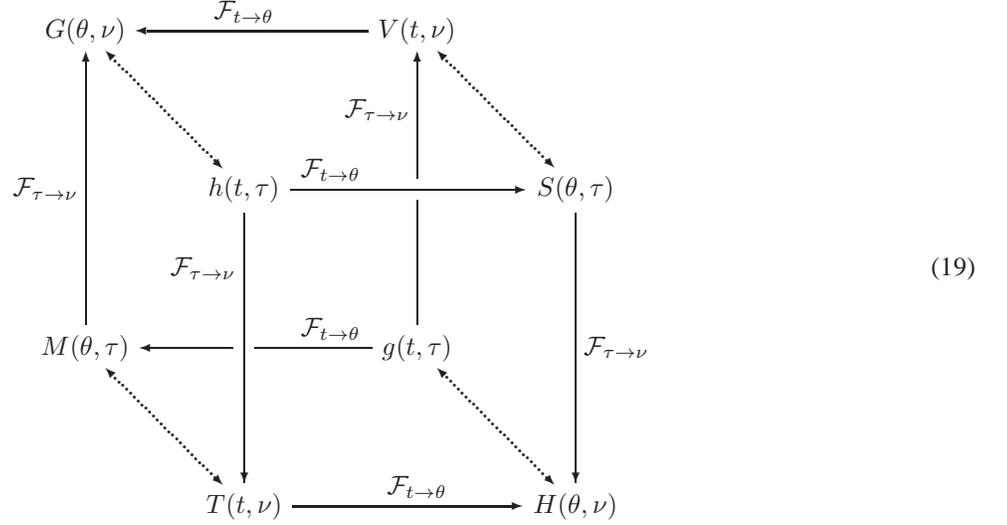
\begin{equation}
\begin{diagram}%[notextflow]
 G(\theta,\nu)   &    &\lTo^{\F_{t\to\theta}} &      &   V(t,\nu)   \\
      & \rdADots &    &      & \uTo^{\F_{\tau\to\nu}}& \rdADots  \\
\uTo^{\F_{\tau\to\nu}} &    &   h(t,\tau)   & \rTo^{\F_{t\to\theta}}  & \HonV   &    &  S(\theta,\tau)  \\
      &    & \dTo^{\F_{\tau\to\nu}}  &      & \vLine  \\
   M(\theta,\tau)  & \lTo & \VonH   & \rLine^{\F_{t\to\theta}} &   g(t,\tau)   &    & \dTo_{\F_{\tau\to\nu}}\\
      & \rdADots &      &      &      & \rdADots  \\
      &    &   T(t,\nu)   &      & \rTo^{\F_{t\to\theta}}  &    &  H(\theta,\nu)  \\
\end{diagram}\label{cube}
\end{equation}
We can derive the following input-output relationships,
\begin{equation}\label{fours}
\begin{aligned}
y(t) &= \int h(t,\tau)x(t-\tau)\dtau   & y(t) &= \doubint S(\theta,\tau)e^{j2\pi\theta t}x(t-\tau)\dtheta\dtau\\
y(t) &= \int T(t,\nu)e^{j2\pi\nu t}X(\nu)\dnu & Y(\theta) &= \int H(\theta-\nu,\nu)X(\nu)\dnu
\end{aligned}
\end{equation}
and
\begin{equation}\label{dualfours}
\begin{aligned}
Y(\theta) &= \int G(\theta,\nu)X(\theta-\nu)\dnu & Y(\theta) &= \doubint V(t,\nu)e^{-j2\pi t\theta}X(\theta-\nu)\dt\dnu\\
Y(\theta) &= \int M(\theta,\tau)e^{-j2\pi\tau\theta}x(\tau)\dtau & y(t) &= \int g(t-\tau,\tau)x(\tau)\dtau
\end{aligned}
\end{equation}
We can relate the eight system functions to the four kernel system functions as follows,
\begin{eqnarray}
k_0(t,\tau) =& h(t,t-\tau) &= g(t-\tau,\tau)\label{impulsed}\\
k_1(\theta,\tau) =& \displaystyle\doubint\!S(\nu,t)e^{j2\pi(t+\tau)(\nu-\theta)}\dnu\dt  &= M(\theta,\tau)e^{-j2\pi\tau\theta}\label{oneSM}\\
k_2(t,\nu) =& T(t,\nu)e^{j2\pi t\nu} &= \doubint\!V(\tau,\theta)e^{j2\pi(t-\tau)(\theta+\nu)}\dtau\dtheta\\
k_3(\theta,\nu) =& H(\theta-\nu,\nu) &= G(\theta,\theta-\nu)\label{k3HG}
\end{eqnarray}
$S(\theta,\tau)$ and $V(t,\nu)$ are distinctive in that their
input-output characterizations and relations to the kernel system
functions involve double integrals. In fact, it is the double integral
formulation involving $S(\theta,\tau)$ in (\ref{fours}) with the interpretation that the output
is a superposition of time-delayed and Doppler-shifted copies of the
input that makes $S(\theta,\tau)$ an extremely useful
characterization. For completeness, we note the inverse relations, 
\begin{eqnarray}
S(\theta,\tau) =& \displaystyle\doubint\!k_1(\nu,t)e^{j2\pi(t+\tau)(\nu-\theta)}\dnu\dt\\
V(t,\nu) =& \displaystyle\doubint\!k_2(\tau,\theta)e^{j2\pi(t-\tau)(\theta+\nu)}\dtau\dtheta
\end{eqnarray}
and note the following relationship between the dual characterizations $h(t,\tau)$ and $G(\theta,\nu)$,
\begin{eqnarray}
h(t,\tau)     &=& \doubint G(\theta,\nu) e^{j2\pi\theta t}e^{-j2\pi(t-\tau)(\theta-\nu)}\dtheta\dnu\\
G(\theta,\nu) &=& \doubint h(t,\tau) e^{-j2\pi t\theta}e^{j2\pi(\theta-\nu)(t-\tau)}\dt\dtau.
\end{eqnarray}

Although less commonly used in the literature, $k_3(\theta,\nu)$ plays a pivotal role in understanding the narrowband and wideband characterizations~\cite{rickardphd}. We note the mapping between $k_3$ and $S$,
\begin{eqnarray}
k_3(\theta,\nu) &=& \int S(\theta-\nu,\tau)e^{-j2\pi\tau\nu}\dtau\label{Stok3}\\
S(\theta,\tau)   &=& \int k_3(\theta+\nu,\nu)e^{j2\pi\tau\nu}\dnu.\label{k3toS}
\end{eqnarray}
which can be derived directly from the input-output channel characterizations.
In the kernel system formulation of the channel, the outputs could be simply expressed in term of the kernel functions for inputs that were impulses in time and frequency. For the above characterizations, these relations are:
\begin{itemize}
\item The response to $\delta(t-t_0)$ is $h(t,t-t_0)$ with spectrum $M(\theta,t_0)e^{-j2\pi\theta t_0}$.
\item The response to $e^{j2\pi\theta_0 t}$ is $T(t,\theta_0)e^{j2\pi t\theta_0}$ with spectrum $H(\theta-\theta_0,\theta_0)$.
\end{itemize}

For clarity, we display just the front face of the cube in (\ref{cube}), which details the Fourier transform relationships among the four most commonly used system functions.
\begin{equation}
\begin{diagram}
h(t,\tau)       & \rTo^{\F_{t\to\theta}} &  S(\theta,\tau)\\
\dTo^{\F_{\tau\to\nu}} &                 & \dTo^{\F_{\tau\to\nu}}\\
T(t,\nu)        & \rTo^{\F_{t\to\theta}} & H(\theta,\nu)
\end{diagram}
\end{equation}

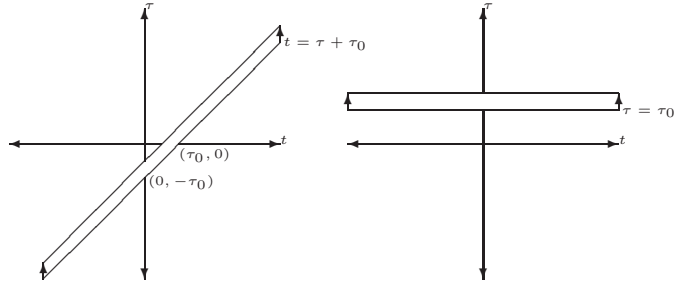
\begin{figure}
\begin{center}
\setlength{\unitlength}{.9cm}
\begin{picture}(9,4)
\put(2,2){\line(1,0){.25}}\put(2.5,2){\vector(1,0){1.5}}%\put(2,2){\vector(1,0){2}}
\put(2,2){\vector(0,1){2}}
\put(2,2){\vector(-1,0){2}}
\put(2,2){\line(0,-1){.25}}\put(2,1.5){\vector(0,-1){1.5}}%\put(2,2){\vector(0,-1){2}}
\put(4,2){\tiny$t$}\put(2,4){\tiny$\tau$}
\put(7,2){\vector(1,0){2}}
\put(7,2){\line(0,1){.5}}\put(7,2.75){\vector(0,1){1.25}}%\put(7,2){\vector(0,1){2}}
\put(7,2){\vector(-1,0){2}}
\put(7,2){\vector(0,-1){2}}
\put(9,2){\tiny$t$}\put(7,4){\tiny$\tau$}

\put(2.5,1.8){\tiny$(\tau_0,0)$}
\put(2.05,1.4){\tiny$(0,-\tau_0)$}
\put(.5,0){\line(1,1){3.5}}
\put(.5,.25){\line(1,1){3.5}}
\put(.5,0){\vector(0,1){.25}}
\put(4,3.5){\vector(0,1){.25}}
\put(4.05,3.45){\tiny$t=\tau+\tau_0$}

\put(5,2.5){\line(1,0){4}}
\put(5,2.75){\line(1,0){4}}
\put(5,2.5){\vector(0,1){.25}}
\put(9,2.5){\vector(0,1){.25}}
\put(9.05,2.45){\tiny$\tau=\tau_0$}
\end{picture}
\caption{$k_0(t,\tau)$ (left) and $h(t,\tau)$ (right) for the delay-by-$\tau_0$ channel.}\label{toosimple}
\end{center}
\end{figure}

\begin{figure}
\begin{center}
\setlength{\unitlength}{.9cm}
\begin{picture}(5,4)
\put(0,2){\vector(1,0){2}}
\put(0,2){\vector(0,1){2}}
\put(0,2){\vector(-1,0){2}}
\put(0,2){\vector(0,-1){2}}
\put(2,2){\tiny$\theta$}\put(0,4){\tiny$\tau$}

\put(1,3){\vector(0,1){.25}}\put(1.075,2.95){\tiny$(\theta_0,\tau_0)$}
\put(1.025,3){\vector(0,1){.25}}

\put(5,2){\vector(1,0){2}}
\put(5,2){\vector(0,1){2}}
\put(5,2){\vector(-1,0){2}}
\put(5,2){\vector(0,-1){2}}
\put(7,2){\tiny$a$}\put(5,4){\tiny$b$}

\put(6,3){\vector(0,1){.25}}\put(6.075,2.95){\tiny$(a_0,b_0)$}
\put(6.025,3){\vector(0,1){.25}}
\end{picture}
\caption{$S(\theta,\tau)$ for one-path channel with delay $\tau_0$ and Doppler shift $\theta_0$ (left); $\Lop(a,b)$ for one-path channel with delay $b_0$ and time dilation $a_0$ (right).}\label{toosimple2}
\end{center}
\end{figure}
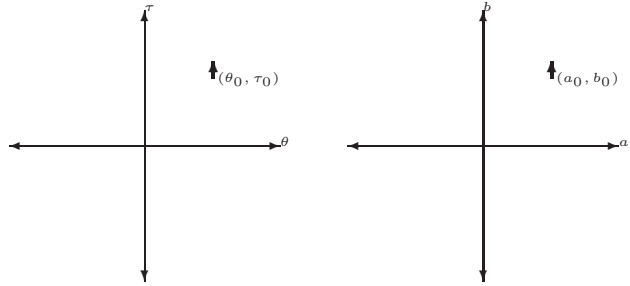

In order to get some intuition concerning the channel characterization functions (both the four kernel functions (\ref{fourks}) and the additional eight characterizations listed in Table~\ref{eight}, we examine simple channels. Consider the time-invariant channel that consists of a pure delay. 
\begin{equation*}
x(t) \to\boxed{\text{channel}}\to x(t-\tau_0)
\end{equation*}
In the case of the time-varying impulse response kernel, this channel
is represented by $k_0(t,\tau)=\delta(t-\tau-\tau_0)$.  In the case of
the time-varying impulse response, this channel is represented by
$h(t,\tau)=\delta(\tau-\tau_0)$.  Plots of these two functions are
displayed in Figure~\ref{toosimple}. One useful attribute of a system
function is for a visual inspection of the function to readily reveal
some physical properties of the channel.  In the case of
Figure~\ref{toosimple}, we see that, for $k_0(t,\tau)$, a diagonal
delta function line crossing through $(0,-\tau_0)$ and $(\tau_0,0)$
arises from a delay of $\tau_0$. For $h(t,\tau)$, a delay of $\tau_0$
corresponds to a horizontal delta function line $\tau_0$ from the
origin. A channel with several reflections (i.e., several different
delays), would thus correspond to a system function with several
parallel delta function lines. When the channel involves both a simple
delay and a Doppler shift, the simple delta function lines for both
$k_0(t,\tau)$ and $h(t,\tau)$ are modulated by the Doppler
shift. Table~\ref{scf} displays the twelve system functions for the delay
and delay-Doppler channels. The system function with the simplest form
is $S(\theta,\tau)$ which is the product of delta
functions. From this, we interpret a region of localized energy in
$S(\theta,\tau)$ centered at $(\theta_0,\tau_0)$ as arising from an
echo path with delay $\tau_0$ and Doppler shift $\theta_0$; see
Figure~\ref{toosimple2}.

\begin{table}
\[
\begin{array}{|c|c|c|}\cline{2-3}
\multicolumn{1}{c|}{}  & y(t)=x(t-\tau_0)                  & y(t)=x(t-\tau_0)e^{j2\pi\theta_0 t}\\\hline
k_0(t,\tau)      & \delta(t-\tau-\tau_0)        & \delta(t-\tau-\tau_0)e^{j2\pi\theta_0 t}\\\hline
k_1(\theta,\tau) & e^{-j2\pi(\tau+\tau_0)\theta}& e^{-j2\pi(\tau+\tau_0)(\theta-\theta_0)}\\\hline
k_2(t,\nu)       & e^{j2\pi (t-\tau_0)\nu}      & e^{j2\pi t(\nu+\theta_0)}e^{-j2\pi \tau_0\nu}\\\hline
k_3(\theta,\nu)  & \delta(\theta-\nu)e^{-j2\pi \tau_0\nu} &  \delta(\theta-\nu-\theta_0)e^{-j2\pi \tau_0\nu}\\\hline\hline
h(t,\tau)        & \delta(\tau-\tau_0)          & \delta(\tau-\tau_0) e^{j2\pi t\theta_0}\\\hline
S(\theta,\tau)   & \delta(\tau-\tau_0)\delta(\theta)  & \mbox{\boldmath{$\delta(\tau-\tau_0)\delta(\theta-\theta_0)$}}\\\hline
T(t,\nu)         & e^{-j2\pi \tau_0\nu}               & e^{j2\pi \theta_0 t}e^{-j2\pi \tau_0\nu} \\\hline
H(\theta,\nu)    & e^{-j2\pi \tau_0\nu}\delta(\theta) & e^{-j2\pi \tau_0\nu}  \delta(\theta-\theta_0)\\\hline
G(\theta,\nu)    & e^{-j2\pi \tau_0\theta}\delta(\nu) & e^{-j2\pi \tau_0(\theta-\theta_0)}\delta(\nu-\theta_0) \\\hline
V(t,\nu)         & \delta(t-\tau_0)\delta(\nu)        &e^{j2\pi \tau_0\theta_0}\delta(t-\tau_0)\delta(\nu-\theta_0)\\\hline
M(\theta,\tau)   & e^{-j2\pi \tau_0\theta}            & e^{-j2\pi \tau_0(\theta-\theta_0)}e^{j2\pi\tau\theta_0}\\\hline
g(t,\tau)        & \delta(t-\tau_0)                   &e^{j2\pi \tau_0\theta_0}\delta(t-\tau_0)e^{j2\pi\tau\theta_0}\\\hline
\end{array}
\]
\caption{Time-frequency characterization functions for the one-path delay and one-path delay-Doppler channels. $S(\theta,\tau)$ has a very simple form for the one-path delay-Doppler channel.}\label{scf}
\end{table}

\subsection{Wideband Characterizations}\label{wc}
Starting from the wideband channel characterization, 
\begin{equation}\label{wbch}
y(t)=\doubint \Lop(a,b)\frac{1}{\sqrt{|a|}}x\left(\frac{t-b}{a}\right)\da\db.
\end{equation}
we derive the frequency domain to frequency domain mapping,
\begin{subequations}
\begin{eqnarray}
Y(\theta)&=&\tripint \Lop(a,b)\frac{1}{\sqrt{|a|}}x\left(\frac{t-b}{a}\right)e^{-j2\pi t\theta}\da\db\dt\\
&=&\tripint \Lop(a,b)\sqrt{|a|}x\left(t^\prime\right)e^{-j2\pi (at^\prime+b)\theta}\da\db\dt^\prime\\
&=&\doubint \Lop(a,b)\sqrt{|a|}X\left(a\theta\right)e^{-j2\pi b\theta}\da\db
\end{eqnarray}
\end{subequations}
and defining,
\begin{equation}
\Lop^{(2)}(a,\theta)=\int \Lop(a,b)e^{-j2\pi b\theta}\db,
\end{equation}
we obtain
\begin{equation}\label{wideff}
Y(\theta)=\int \Lop^{(2)}(a,\theta)\sqrt{|a|}X\left(a\theta\right)\da
\end{equation}

Table~\ref{scf_ts} displays the wideband characterization functions for the one-path delay and one-path delay-dilation channels. In the narrowband case, $S(\theta,\tau)$ is the product of delta functions for the one-path delay-Doppler channel; In the wideband case, the one-path delay-dilation channel is the product of delta functions. We interpret a region of concentrated energy in $\Lop(a,b)$ centered at $(a_0,b_0)$ as arising from an echo path with delay $b_0$ and dilation parameter $a_0$.

\begin{table}
\[
\begin{array}{|c|c|c|}\cline{2-3}
\multicolumn{1}{c|}{}  & y(t)=x(t-b_0)                  & y(t)=\frac{1}{\sqrt{|a_0|}}x(\frac{t-b_0}{a_0})\\\hline
\Lop(a,b)            & \delta(a-1)\delta(b-b_0) & \delta(a-a_0)\delta(b-b_0)\\\hline
\Lop^{(2)}(a,\theta) & \delta(a-1)e^{-j2\pi b_0\theta} & \delta(a-a_0)e^{-j2\pi b_0\theta}\\\hline
\end{array}
\]
\caption{Time-scale characterization functions for the one-path delay and one-path delay-dilation channels.}\label{scf_ts}
\end{table}

\subsection{Narrowband and Wideband Correspondence}\label{narrowtowidec}
In this section we briefly examine the correspondence between the narrowband and wideband channel models.
More specifically, we wish to link the narrowband channel model characterized by the dozen system functions discussed above, one of which was described by the time-frequency integral operator,
\begin{equation}\label{tfop}
\Nop_S x (t) \defeq \doubint S(\theta,\tau)x(t-\tau)e^{j2\pi\theta t}\dtau\dtheta
\end{equation}
to the wideband channel description embodied in the time-scale integral operator,
\begin{equation}\label{tsop}
\Wop_\Lop x (t)\defeq\doubint\Lop(a,b)\frac{1}{\sqrt{|a|}}x\left(\frac{t-b}{a}\right)\da\db. 
\end{equation}
We are interested in the mapping between $S$ and $\Lop$ for
$\Nop_S=\Wop_\Lop$. The approach taken here differs from the
traditional interpretation of the narrowband characterization as an
approximation of the wideband characterization when applied to
narrowband signals. This approximation is discussed in detail in, for
example,~\cite{fowler91,chaiyasena92,sibul94,weiss94,sibul97,rebollo-neira97,sibul98}.
We do not consider the narrowband description of the channel as an
approximation of the wideband channel, but rather look at the two
descriptions without constraining the properties of the input signal.

We first establish the relation from wideband to narrowband, showing that for every time-scale kernel, there exists a corresponding time-frequency kernel. Starting from (\ref{tsop}), we have
\begin{subequations}
\begin{eqnarray}
y(t)&=&\doubint\Lop(a,b)\frac{1}{\sqrt{|a|}}x\left(\frac{t-b}{a}\right)\da\db\\
&=&\int\left(\int\sqrt{|a|}\Lop(a,t-a\tau)\da\right)x(\tau)\dtau,
\end{eqnarray}
\end{subequations}
and therefore,
\begin{equation}\label{ts2tf}
k_0(t,\tau) = \int \sqrt{|a|}\Lop(a,t-a\tau)\da.
\end{equation}

Returning to the mapping from $\Lop(a,b)$ to the narrowband characterizations, starting from (\ref{ts2tf}), the remaining system functions can be related to $\Lop(a,b)$ as follows,
\begin{equation}
h(t,\tau)  = \int \sqrt{|a|}\Lop(a,(1-a)t+a\tau)\da\label{htoL}
\end{equation}
and, taking the Fourier transform of (\ref{htoL}) with respect to $t$, we obtain,
\begin{equation}
\boxed{S(\theta,\tau) = \doubint \sqrt{|a|}\Lop(a,(1-a)t+a\tau)e^{-j2\pi\theta t}\da\dt\label{SfromL}}
\end{equation}
Using (\ref{ts2tf}), it is possible to relate $\Lop$ to all twelve narrowband representations~\cite{rickardphd}. 

It is also possible to express  $\Lop(a,b)$ in terms of $S(\theta,\tau)$,
\begin{equation}\label{LfromS}
\boxed{\Lop(a,b)= \doubint \frac{|\theta|}{\sqrt{|a|}}S((1-a)\theta,\tau)e^{j2\pi\theta(b-a\tau)}\dtheta\dtau}
\end{equation}
although the mapping relies on the assumption that the input signal
has no DC component; see~\cite{rickardphd} for a discussion of this
mapping. We can observe from (\ref{wideff}) that, in the wideband
model, the DC input component can only affect the DC output
component. Intuitively, it is clear that rescaling the time axis and
shifting in time a DC signal does not have any effect, and all the
time-scale channel can do is amplify or attenuate the DC component of
a signal. This is not the case in the narrowband model. For example,
from $k_3(\theta,\nu)$ in (\ref{fourks}) it is clear that the DC input
signal component can affect any output frequency component. Therefore,
there are time-frequency characterizations which have no corresponding
time-scale representation.

We look to some simple channel models and examine the mappings between $\Lop$ and $S$.
We first consider the wideband (delay-dilation) single path channel,
\begin{equation}
\Lop(a,b) = \delta(a-a_0)\delta(b-b_0).\label{singleL}
\end{equation}
It follows from (\ref{SfromL}) that,
\begin{equation}\label{singleSfromL}
S(\theta,\tau) =  
\left\{ \begin{array}{c@{\quad:\quad}c} \frac{\sqrt{|a_0|}}{|1-a_0|}e^{-j2\pi\theta\frac{b_0-a_0\tau}{1-a_0}} & a_0\neq 1 \\ 
                                        \delta(\theta)\delta(\tau-b_0) & a_0 = 1\end{array}\right.
\end{equation}
and, substituting this into (\ref{nc}) we obtain $y(t)=x_{a_0,b_0}(t)$, as expected.

We can derive the time-varying impulse response characterization $h(t,\tau)$ for the wideband single path channel,
\begin{subequations}
\begin{eqnarray}
h(t,\tau) &=& \int S(\theta,\tau)e^{j2\pi\theta t}\dtheta\\
         &=& \int \frac{\sqrt{|a_0|}}{|1-a_0|}e^{-j2\pi\theta\frac{b_0-a_0\tau}{1-a_0}}e^{j2\pi\theta t}\dtheta\\
         &=& \frac{\sqrt{|a_0|}}{|1-a_0|}\delta\left(\frac{b_0-a_0\tau-(1-a_0)t}{1-a_0}\right)\\
         &=& \sqrt{|a_0|}\delta(b_0-a_0\tau-(1-a_0)t)
\end{eqnarray}
\end{subequations}
which is also valid when $a_0=1$. We can compare this result to that of the single narrowband path (delay by $\tau_0$, Doppler shift by $\theta_0$) channel, $h(t,\tau)=\delta(\tau-\tau_0)e^{j2\pi t\theta_0}$. The wideband path gives rise to a delta function line with slope $\frac{a_0-1}{a_0}$ intersecting the $\tau$-axis at $b_0/a_0$; The narrowband path gives rise to a modulated delta function line parallel to the $t$-axis intersecting the $\tau$-axis at $\tau_0$.

We now turn to the expression of the narrowband (delay-Doppler) single path in the wideband model:
\begin{equation}
S(\theta,\tau) = \delta(\theta-\theta_0)\delta(\tau-\tau_0)\label{singleS}
\end{equation}
If we ignore the difficulties arising from the instabilities on the $a=1$ line~\cite{rickardphd}, it follows from (\ref{LfromS}) that
\begin{equation}\label{singleLfromS}
\Lop(a,b) = \frac{|\theta_0|}{\sqrt{|a|}(1-a)^2}e^{j2\pi\theta_0\frac{b-a\tau_0}{1-a}} 
\end{equation}
and, substituting this into (\ref{firstL}), we indeed obtain, $y(t)= x_{\tau_0,\theta_0}(t)$.

The various channel characterizations for the simple one-path models
(including the time-invariant one-path model) are displayed in
Table~\ref{bigtab}.  We note that the one-path delay-dilation channel
requires infinite support in time-frequency (\ref{singleSfromL})
whereas it requires only point support in time-scale
(\ref{singleL}). On the other hand, the one-path delay-Doppler channel
requires infinite support in time-scale (\ref{singleLfromS}) whereas
it requires only point support in time-frequency. Thus, since we are
interested in channels which have finite support in time-frequency or
time-scale (as we will see in the next sections), the choice of
channel model is crucial and must be appropriate to the signaling
environment (i.e., narrowband or wideband). Examination of
$k_3(\theta,\nu)$ for the one-path channels reveals that it is
possible (up to a scaling constant) for the one-path delay-Doppler and
the one-path delay-dilation channels to have the same effect on a
narrowband signal (eg, $X(\nu)=\delta(\nu-\nu_0)$) by setting
$\tau_0=b_0/a_0$ and $\theta_0=v_0\left(\frac{a_0-1}{a_0}\right)$.

\begin{table}
\begin{center}
\[
\begin{array}{|c|c|c|c|}\cline{2-4}
\multicolumn{1}{c|}{} & \text{one-path delay only} & \text{one-path delay-Doppler, $\theta_0\neq 0$} & \text{one-path delay-dilation, $a_0\neq 1$} \\\hline
S(\theta,\tau)   & \delta(\theta)\delta(\tau-t_0) &\mbox{\boldmath{$\delta(\theta-\theta_0)\delta(\tau-\tau_0)$}} & \frac{\sqrt{|a_0|}}{|1-a_0|}e^{-j2\pi\theta\frac{b_0-a_0\tau}{1-a_0}}\\\hline
\Lop(a,b)        & \delta(a-1)\delta(b-t_0) &  \frac{|\theta_0|}{\sqrt{|a|}(1-a)^2}e^{j2\pi\theta_0\frac{b-a\tau_0}{1-a}}
& \mbox{\boldmath{$\delta(a-a_0)\delta(b-b_0)$}} \\\hline
h(t,\tau)        & \delta(\tau-t_0) & \delta(\tau-\tau_0)e^{j2\pi t\theta_0}          & \sqrt{|a_0|}\delta((1-a_0)t+a_0\tau-b_0) \\\hline
k_3(\theta,\nu)  & \delta(\theta-\nu)e^{-j2\pi t_0\nu} & \delta(\theta-\nu-\theta_0)e^{-j2\pi \tau_0\nu} & \sqrt{|a_0|}e^{-j2\pi\theta b_0}\delta(\nu-a_0\theta) \\\hline
\end{array}
\]
\caption{Time-frequency and time-scale characterizations for the one-path delay-Doppler and one-path delay-dilation channels.}\label{bigtab}
\end{center}
\end{table}

\section{Discrete Canonical Channel Models}\label{tfrake}
In this section we develop a general technique for the generation of
canonical channel models and demonstrate the application of the
technique to time-frequency and time-scale kernel operators. 

\subsection{The canonical rake receiver model} 
We begin with the derivation of the canonical model associated with the standard rake receiver.
The classic expression of the sampling theorem for a signal $X(\nu)$ with support $(-W/2,W/2)$ is
\begin{equation}
x(t) = \sum_{n=-\infty}^{\infty}x\left(\frac{n}{W}\right)\frac{\sin\left(\pi W\left(t-\frac{n}{W}\right)\right)}{\pi W\left(t-\frac{n}{W}\right)}.
\end{equation}
An alternative formulation of the sampling theorem~\cite{vantrees71} is obtained by defining $g(t) = x(\alpha-t)$, 
\begin{equation}
g(t) = \sum_{n=-\infty}^{\infty}g\left(\frac{n}{W}\right)\frac{\sin\left(\pi W\left(t-\frac{n}{W}\right)\right)}{\pi W\left(t-\frac{n}{W}\right)}\\
\end{equation}
and thus,
\begin{equation}
x(\alpha-t) = \sum_{n=-\infty}^{\infty}x\left(\alpha-\frac{n}{W}\right)\frac{\sin\left(\pi W\left(t-\frac{n}{W}\right)\right)}{\pi W\left(t-\frac{n}{W}\right)}
\end{equation}
Mapping $(\alpha,t) \rightarrow (t,\tau)$, we obtain,
\begin{equation}\label{altsamp}
x(t-\tau) = \sum_{n=-\infty}^{\infty}x\left(t-\frac{n}{W}\right)\frac{\sin\left(\pi W\left(\tau-\frac{n}{W}\right)\right)}{\pi W\left(\tau-\frac{n}{W}\right)}.
\end{equation}
Following~\cite{vantrees71}, substituting (\ref{altsamp}) into the time-varying impulse response channel characterization (\ref{tvlc}), we obtain
\begin{subequations}
\begin{eqnarray}
y(t) \qeq\int h(t,\tau)x(t-\tau)\dtau\\
\qeq \sum_{n=-\infty}^{\infty}x\left(t-\frac{n}{W}\right)\underbrace{\left[\int h(t,\tau)\frac{\sin\left(\pi W\left(\tau-\frac{n}{W}\right)\right)}{\pi W\left(\tau-\frac{n}{W}\right)}\dtau\right]}_{=h_n(t)}\label{rake1}\\
&\approx& \sum_{n=0}^{L\defeq\lceil T_m/W\rceil}x\left(t-\frac{n}{W}\right)h_n(t)\label{tapline}
\end{eqnarray}
\end{subequations}
where the approximation is made based on the assumption that the channel is causal and has finite multipath spread, $T_m$. That is, $h(t,\tau)= 0, \forall \tau<0, \tau>T_m$. Under this assumption, the approximation (\ref{tapline}) corresponds to $h_n(t)$ for which the mainlobe of the sinc function overlaps with the support of the time-varying impulse response. The tapped-delay line in (\ref{tapline}) forms the basis for the classic rake receiver, where $h_n(t)$'s are usually assumed to be independent of each other.

\subsection{The canonical time-frequency model} 
We now proceed to examine the time-frequency canonical channel model
which was originally derived in~\cite{sayeed99joint}. Alternative, but
similar models are explored in~\cite{giannakis98,thomas00,ma02}. The
path we take in this derivation is essentially the same as that
in~\cite{sayeed99joint}. We look at only the $(0,T)$ portion of the
received waveform, that is, $y(t)1_{(0,T)}(t)$.  Starting from
(\ref{rake1}), we impose the $(0,T)$ restriction and obtain
\begin{equation}
y(t)1_{(0,T)}(t) =
\sum_{n=-\infty}^{\infty}x\left(t-\frac{n}{W}\right)\left[\int
h(t,\tau)1_{(0,T)}(t)\sinc\left(W\left(\tau-\frac{n}{W}\right)\right)\dtau\right]\label{rake1ass}
\end{equation}
Now we expand the $h(t,\tau)1_{(0,T)}(t)$ term as a Fourier series, 
\begin{subequations}
\begin{eqnarray}
h(t,\tau)1_{(0,T)}(t)&=&\sum_{k=-\infty}^{\infty}\frac{1}{T}\left[\int_{0}^{T}h(t',\tau)e^{-j2\pi kt'/T}\dt'\right]e^{j2\pi kt/T}\\
&=& \sum_{k=-\infty}^{\infty}\frac{1}{T}\underbrace{\left[\int_{-\infty}^{\infty}h(t',\tau)1_{(0,T)}(t')e^{-j2\pi kt'/T}\dt'\right]}_{\int_{-\infty}^{\infty}S(\theta,\tau)T\sinc\left(\left(\frac{k}{T}-\theta\right)T\right) e^{-j\pi(k-T\theta)}\dtheta}e^{j2\pi kt/T}\label{fexp}
\end{eqnarray}
\end{subequations}
which is valid for $t\in(0,T)$.

Substituting (\ref{fexp}) into (\ref{rake1ass}) we obtain,
\begin{equation}\label{tfrakeinf}
y(t)  = \sum_{n=-\infty}^{\infty}\sum_{k=-\infty}^{\infty}x\left(t-\frac{n}{W}\right)e^{j2\pi kt/T}\hat{S}\left(\frac{k}{T},\frac{n}{W}\right)
\end{equation}
where, 
\begin{equation}
\hat{S}(\theta,\tau)\defeq
\doubint S(\theta^\prime,\tau^\prime)\sinc\left(\left(\tau-\tau^\prime\right)W\right)\sinc\left(\left(\theta-\theta^\prime\right)T\right) e^{-j\pi(\theta-\theta^\prime)T}\dtheta'\dtau'\label{smoothS}
\end{equation}
(\ref{tfrakeinf}) is valid for that part of any bandlimited signal received during $(0,T)$.

Under the path scatterer interpretation we assume that the channel
introduces a maximum delay spread of $T_m$ and maximum Doppler spread
of $B_d$, that is, $S(\theta,\tau)$ has support in
$(-B_d,B_d)\times(0,T_m)$. In the smoothed version of
$S(\theta,\tau)$ in (\ref{smoothS}), if we consider only the terms in
(\ref{tfrakeinf}) where the main lobe of the smoothing kernel (which
has size $(-1/T,1/T)$-by-$(-1/W,1/W)$) overlaps with the support of
$S(\theta,\tau)$, we need only sum over $n=0,\ldots,N$ where $N=\lceil
WT_m \rceil$ and $k=-K,\ldots,K$ where $K=\lceil TB_d \rceil$. We thus obtain the canonical representation of the time-frequency channel model, 
\begin{equation}\label{cantfrake}
y(t)  = \sum_{n=0}^{\lceil WT_m \rceil}\sum_{k=-\lceil TB_d \rceil}^{\lceil TB_d \rceil}x\left(t-\frac{n}{W}\right)e^{j2\pi kt/T}\hat{S}\left(\frac{k}{T},\frac{n}{W}\right)
\end{equation}

\subsection{Restatement}\label{rest}

The double sum time-frequency channel formulation (\ref{tfrakeinf}) was obtained by assuming,
\begin{itemize}
\item the input signal is bandpass with bandwidth $W$, and
\item the output signal is analyzed only for $t\in(0,T)$.
\end{itemize}
With these assumptions in mind, we define the following two projection operators,
\begin{equation}
P_Tx(t) \defeq 1_{[0,T]}(t)x(t)
\end{equation}
and,
\begin{equation}
Q_Wx(t) \defeq \F^{-1}\{1_{[-W/2,W/2]}(\omega)\F\{x(t)\}(\omega)\},
\end{equation}
and using the following two operators, the translation operator,
\begin{equation}
T_\tau x(t) \defeq x(t-\tau),
\end{equation}
and the modulation operator,
\begin{equation}
M_\nu x(t) \defeq x(t)e^{j2\pi\nu t},
\end{equation}
we can rewrite (\ref{tfrakeinf}) as,
\begin{equation}
P_T\Nop_SQ_W = \sum_{m,n}c_{m,n}P_TM_{\frac{1}{T}}^mT_{\frac{1}{W}}^nQ_W
\end{equation}
where the $c_{m,n} = \hat{S}\left(\frac{m}{T},\frac{n}{W}\right)$ and
$\Nop_S$ is the narrowband channel operator defined in (\ref{tfop}).
Restating the channel operator in this setting, we can ask what
general properties of the operators allow us to express the channel as
a double summation of transformed input waveforms. In this section, we
determine properties of the operators that are sufficient conditions
for the existence of such an expansion. Our goal is to develop an
analogous time-scale canonical channel model.  That is, in
Section~\ref{tscm} we propose projections $P$ and $Q$ such that,
\begin{equation}
P\Wop_LQ = \sum_{m,n}c_{m,n}PD_{a_0}^mT_{b_0}^nQ
\end{equation}
for some choice of dilation and translation spacing parameters ($a_0$
and $b_0$), where the $c_{m,n}$ depend on $\Lop$, and $D$ is the
dilation operator,
\begin{equation}
D_ax(t) \defeq \frac{1}{\sqrt{|a|}}x\left(\frac{t}{a}\right),
\end{equation}
for the wideband channel operator defined in (\ref{tsop}).

\subsection{Generalization}\label{gen}
For the statement of the general theorem, we require the following definition.
\begin{definition}[paired-up operators]
$P$ and $U$ are paired-up operators with generator $e_0$ iff,
\begin{enumerate}
\item $P$ is an orthogonal projection in $L^2(\R)$
\item $U$ is unitary in $L^2(\R)$ 
\item $PU=UP$
\item $\exists e_0\in\Ran P \sut \{U^me_0\,:\,m\in\Z\}$ is an orthonormal basis for $\Ran P$
\end{enumerate}
\end{definition}

Using two different pairs of paired-up operators, the following theorem gives a sufficient condition for the channel expansion.
\begin{theorem}
If $(P,U)$ and $(Q,V)$ are both paired-up operators with generator elements $e_0$ and $f_0$ respectively, $H$ is a bounded operator, and $\exists c_{m,n}$ such that 
\begin{equation}
\sum_{m,n} c_{m,n}\Langle V^{n+k}f_0,U^{l-m}e_0\Rangle=\Langle H V^kf_0,U^le_0\Rangle,\quad\forall k,l,\label{solvethis}
\end{equation}
then, 
\begin{equation}
PHQ = \sum_{m,n} c_{m,n}PU^mV^nQ\label{gthm}
\end{equation}
\end{theorem}
The proof of this theorem can be found in Appendix~\ref{mainp} and a method for calculating the coefficients $c_{m,n}$ can be found in Appendix~\ref{ccalc}.

\subsection{Revisiting time-frequency}\label{rtf}
The example we have seen so far of the application of this theorem corresponds to the situation
\begin{itemize}
\item $(P,U,e_0)=(P_T,M_{\frac{1}{T}},\frac{1}{\sqrt{T}}1_{[0,T]}(t))$
\item $(Q,V,f_0)=(Q_W,T_{\frac{1}{W}},\sqrt{W}\sinc(W t))$
\end{itemize}
for the operator $H=\Nop_S$ of the form, 
\begin{equation}
 Hx(t) = \doubint S(\theta,\tau)e^{j2\pi\theta t}x(t-\tau)\dtheta\dtau.
\end{equation}
Modulation and translation operators are a natural fit with our channel description, $\Nop_S$, which describes the channel as a (continuous) summation of time and frequency shifts of the input signal. In Appendix~\ref{ccalc} we demonstrate the coefficient calculation procedure for these specific operators. The procedure correctly derives the result $c_{m,n}=\hat{S}\left(\frac{m}{T},\frac{n}{W}\right)$ where $\hat{S}$ is defined in (\ref{smoothS}).

\subsection{Time-scale canonical model}\label{tscm}
We now develop the time-scale canonical characterization. For other
possible extensions to time-scale, see the approach
in~\cite{doroslovacki96,zhang00,zhang01} using wavelet packet
modulation. 

The Mellin transform (also known as the scale transform) of a signal $x\in L^2(0,\infty)$ is defined by
\begin{equation}
\M x(\omega)\defeq\int_0^{\infty}e^{-j2\pi\omega \,\ln\,t}x(t)\frac{\dt}{\sqrt{t}}
\end{equation}
which represents the composition of two unitary transformations
\[ x(t)  \rightarrow e^{t/2}x(e^t)  \stackrel{\F_{t\rightarrow \omega}}{\longrightarrow} \M x(\omega).\]
For more information on the Mellin transform and its use in time-frequency analysis we refer the reader to~\cite{cohen95time-frequency}. For the time-scale canonical characterization, we will require the projection operator in the Mellin transform domain
\begin{equation}
R_\alpha \defeq \M^{-1} 1_{[-\alpha/2,\alpha/2]}\M
\end{equation}
which acts on a function $x\in L^2(0,\infty)$ as follows
\[ x(t) \stackrel{\M}{\longrightarrow} \M x(\omega) \stackrel{1_{[-\alpha/2,\alpha/2]}}{\longrightarrow}  1_{[-\alpha/2,\alpha/2]}(\omega)\M x(\omega) \stackrel{\M^{-1}}{\longrightarrow} R_\alpha x(t)\]
where $\alpha>0$ defines the cut-off Mellin ``frequency''. Explicitly, this means
\begin{equation}
\label{eq:Q}
R_\alpha x(t) = \int_0^{\infty}\frac{1}{\sqrt{t\tau}}\sinc[\alpha(\ln\,t-\ln\,\tau)]x(\tau)\dtau,
~~t>0.
\end{equation}
Using the characteristic function in the Mellin transform domain,
\begin{equation}
\Gamma_0(\omega) = 1_{\left[-\frac{1}{2\ln a_0},\frac{1}{2\ln a_0}\right]}(\omega),
\end{equation}
leads to the scale generator
\begin{equation}
\gamma_0(t) = \left\{ \begin{array}{r@{\quad:\quad}l} \frac{1}{\sqrt{\ln a_0}}\frac{1}{\sqrt{t}}\sinc\left(\frac{\ln t}{\ln a_0}\right) & t>0\\ 0\hspace{.4in}& t<0\end{array}\right.\label{e0}
\end{equation}
For further details on the Mellin transform domain and its generators, consult~\cite{sundaram97}. 

It can be shown that $(P,U,e_0)=(R_{\frac{1}{2\ln a_0}},D_{a_0},\gamma_0(t))$ are paired-up, and thus for the time-scale model, we use the following paired-up operators,
\begin{itemize}
\item $(P,U,e_0)=\left(R_{\frac{1}{2\ln a_0}},D_{a_0},\frac{1}{\sqrt{\ln a_0}}\frac{1}{\sqrt{t}}\sinc\left(\frac{\ln t}{\ln a_0}\right)\right)$
\item $(Q,V,f_0)=(Q_{\frac{1}{b_0}},T_{b_0},\frac{1}{\sqrt{b_0}}\sinc(\frac{t}{b_0}))$
\end{itemize}
to decompose the wideband channel corresponding to the 
operator $H=\Wop_\Lop$ of the form, 
\begin{equation}\label{Hwide}
Hx(t) = \doubint \Lop(a,b)\frac{1}{\sqrt{|a|}}x\left(\frac{t-b}{a}\right)\da\db
\end{equation}
into a discrete double summation,
\begin{equation}\label{ddsDT}
P\Wop_LQ = \sum_{m,n}c_{m,n}PD_{a_0}^mT_{b_0}^nQ.
\end{equation}
In Appendix~\ref{ccalc} we calculate the coefficients in the time-scale case,
\begin{equation}
c_{m,n} = \doubint\Lop(a,b)\sinc\left(m-\frac{\ln a}{\ln a_0}\right)\sinc\left(n-\frac{b}{ab_0}\right)\da\db,
\end{equation}
and the canonical time-scale model is then\footnote{It is also possible to reverse the order of application of $T$ and $D$ in (\ref{ddsDT}). In such case, the generated model in fact is more like that in (\ref{Hwide}) in that the time-shifts are not scaled.}
\begin{equation}\label{cts}
y(t) = \sum_{m,n} \frac{c_{m,n}}{a_0^{m/2}}x\left(\frac{t-nb_0a_0^m}{a_0^m}\right).
\end{equation}

\section{Physical Interpretation of Canonical Models}\label{interpsec}
In~\cite{jiang05} the canonical model (\ref{cts}) is obtained by using
two sampling results: the classical Shannon sampling formula for
bandlimited functions, and a similar sampling result for functions
that have finite support in the Mellin transform domain. With the help
of these two formulas,~\cite{jiang05} obtained a decomposition of the
received signal into a series of such as (\ref{cts}) where parameters
$a_0,b_0$ are directly related to transmit signal bandwidth and
transmit signal Mellin domain bandwidth.  The trouble with such a
model is that there are no signals that are perfectly (Fourier)
frequency bandlimited and Mellin transform bandlimited (similar to the
classical result that there are no time-frequency bandlimited signals
except for the trivial zero signal). 
One can argue that the
transmit signal is essentially frequency bandlimited, as well as,
Mellin domain bandlimited, and thus a decomposition of type
(\ref{cts}) should hold approximately. Furthermore, for a practical
application, the infinite series (\ref{cts}) is truncated to a finite
number of terms consistent with the finite size of the wideband
spreading function $\lc$. Thus, another approximation is introduced,
so overall one might expect that not much is lost by the initial
assumption of joint Fourier frequency - Mellin domain band
limitedness.

In contrast, the approach we took in~\cite{balan04} does not suffer
from the shortcomings outlined above. This different approach uses all
the three players: the sender, the channel, and the receiver. The
sender prepares the transmit signal by tailoring some of its
properties. That is, the signal is embedded into the range of an
orthogonal projection $Q$ (e.g. $Q$ can be an ideal lowpass filter); The
channel acts via the operator $H$ (\ref{eq:1}); and the receiver
observes the channel output but in the observation process applies its
own projection operator $P$ through measurement, e.g. $P$ is a time
cut-off operator. Thus, the entire transmitter-channel-receiver chain is
modeled by a ``sandwich'' of operators $PHQ$ where $P$ and $Q$ are
under the user's control, and $H$ is the channel operator. To simplify
notation, the transmit signal is assumed to lie already in the range of $Q$,
and thus $Q$ often disappears from formulae.  

\begin{figure}[htb]
\begin{center}
\setlength{\unitlength}{3947sp}%
\begingroup\makeatletter\ifx\SetFigFont\undefined%
\gdef\SetFigFont#1#2#3#4#5{%
  \reset@font\fontsize{#1}{#2pt}%
  \fontfamily{#3}\fontseries{#4}\fontshape{#5}%
  \selectfont}%
\fi\endgroup%
\begin{picture}(4782,1524)(64,-748)
\put(1801,-286){\makebox(0,0)[lb]{\smash{{\SetFigFont{12}{14.4}{\rmdefault}{\mddefault}{\updefault}{\color[rgb]{0,0,0}Physical Channel}%
}}}}
\thinlines
{\color[rgb]{0,0,0}\put(4317,-12){\vector( 1, 0){465}}
}%
{\color[rgb]{0,0,0}\put(3438,-736){\framebox(1396,1448){}}
}%
\put(3955,-64){\makebox(0,0)[lb]{\smash{{\SetFigFont{12}{14.4}{\rmdefault}{\mddefault}{\updefault}{\color[rgb]{0,0,0}$P$}%
}}}}
\put(4472, 92){\makebox(0,0)[lb]{\smash{{\SetFigFont{12}{14.4}{\rmdefault}{\mddefault}{\updefault}{\color[rgb]{0,0,0}$y$}%
}}}}
\put(3903,-684){\makebox(0,0)[lb]{\smash{{\SetFigFont{12}{14.4}{\rmdefault}{\mddefault}{\updefault}{\color[rgb]{0,0,0}Receiver}%
}}}}
{\color[rgb]{0,0,0}\put(697,-322){\framebox(672,931){}}
}%
{\color[rgb]{0,0,0}\put( 76,-684){\framebox(1396,1448){}}
}%
{\color[rgb]{0,0,0}\put(231, 92){\vector( 1, 0){466}}
}%
{\color[rgb]{0,0,0}\put(1369, 92){\vector( 1, 0){282}}
}%
{\color[rgb]{0,0,0}\put(3301, 89){\vector( 1, 0){344}}
}%
{\color[rgb]{0,0,0}\put(1651,-586){\framebox(1650,1195){}}
}%
\put(955, 92){\makebox(0,0)[lb]{\smash{{\SetFigFont{12}{14.4}{\rmdefault}{\mddefault}{\updefault}{\color[rgb]{0,0,0}$Q$}%
}}}}
\put(386,143){\makebox(0,0)[lb]{\smash{{\SetFigFont{12}{14.4}{\rmdefault}{\mddefault}{\updefault}{\color[rgb]{0,0,0}$x$}%
}}}}
\put(335,-581){\makebox(0,0)[lb]{\smash{{\SetFigFont{12}{14.4}{\rmdefault}{\mddefault}{\updefault}{\color[rgb]{0,0,0}Transmitter}%
}}}}
\put(2300, 40){\makebox(0,0)[lb]{\smash{{\SetFigFont{12}{14.4}{\rmdefault}{\mddefault}{\updefault}{\color[rgb]{0,0,0}$H$}%
}}}}
{\color[rgb]{0,0,0}\put(3645,-426){\framebox(672,931){}}
}%
\end{picture}%
\caption{Our basic model for a communication channel. \label{fig1}}
\end{center}
\end{figure}
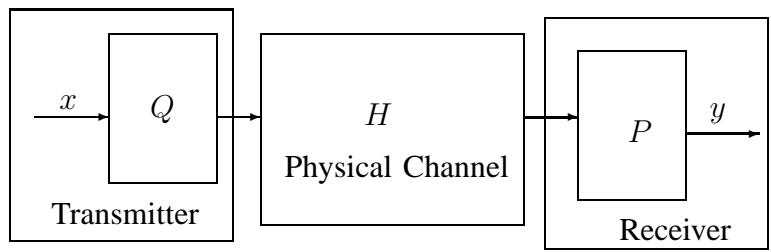

Thus, our basic model for transmitter-to-receiver communication system
contains three blocks (see Figure \ref{fig1}):
\begin{enumerate}
\item A transmit signal shaper, which is mathematically translated into a 
projection $Q$; this can be thought of as the last stage of a modulator
which, for narrowband communication channels, is either a bandpass filter 
around the carrier frequency, or a lowpass filter when analysis is done in the
base band;
\item A physical channel, mathematically modeled by a linear time-varying
system $H$; as such it can be written as in (\ref{eq:1});
\item A received signal observation, which again is translated into another
projection $P$ at the receiver; typically for memoryless source and channels,
this is a time cut-off operator, due to real-time operation constraints.
\end{enumerate}
By changing the transmitter shaping and receiver observation
projections, we obtain the different canonical representations. With
this interpretation in mind, we can revisit previous models.

The standard rake receiver uses a channel model of type:
\[ y(t) = \sum_n h_n(t) x\left(t-\frac{n}{W}\right) \]
which is obtained for $P$ equal to the identity operator (i.e. the
entire channel output is available for processing) and for $Q$ equal
to the projection onto the space of frequency-bandlimited
functions. As mentioned before, $x$ is already assumed to be a frequency-bandlimited signal, thus $x\in \Ran\, Q$.

The time-frequency channel model of~\cite{sayeed99joint} uses the
model (\ref{intro-tfdiscrete}) which is obtained when $P$ is the time
cut-off multiplication by $1_{[0,T]}$ and $Q$ is the ideal lowpass
filter. 

The time-scale channel model of~\cite{balan04} in (\ref{cts}) uses the
ideal Mellin domain lowpass filter as $P$ and the ideal lowpass filter
as $Q$. In other words, the channel output is observed through a scale
filter defined using the Mellin transform. We now present another
canonical model in which the pair of projectors consists of the time
cut-off $1_{[T_1,T_2]}(t)$ for $P$ and the ideal Mellin domain lowpass
filter for $Q$.

\subsection{The Frequency-Scale Canonical Model}\label{sfm}
We now consider a frequency-scale canonical channel characterization
based on the translation operators in frequency and scale.  In the
frequency-scale model, we restrict ourselves to $x(t)$ defined for
$t>0$ and use for the transmitter projection $Q$ the Mellin domain
band limiter $R_\frac{1}{2\ln a_0}$. Thus, the transmitter transmits
scale-limited waveforms. We use for the receiver projection $P$ simply
a time cut-off
\begin{equation}
\label{eq:P}
P_{[T_1,T_2]}x(t)\defeq 1_{[T_1,T_2]}(t)x(t)
\end{equation}
where $T_2>T_1>0$ define the receiver observation time horizon.
The overall chain of operators then decouples into the series
\begin{equation} \label{eq.79} 
PHQ \longrightarrow \sum_{m,n}c_{m,n}P_{[T_1,T_2]}M^m_{1/(T_2-T_1)}D^n_{a_0}R_{\frac{1}{2\ln a_0}}.
\end{equation}
For this model the following theorem gives a decomposition into a 
series of dilated and frequency shifted versions of the input signal.
\begin{theorem}[The Canonical Frequency-Scale Channel Model] Assume
a time-varying channel $H$ defined by (\ref{eq:1}). Then for any
signal $x$ that is Mellin domain bandlimited to $[-\frac{1}{2\ln a_0},
\frac{1}{2\ln a_0}]$, i.e. $x\in \Ran\,Q$,
\begin{equation}
\label{eq:FSmodel}
y(t)\defeq Hx(t)=\sum_{m,n\in\Z} c_{m,n}e^{j2\pi m\Omega t}\frac{1}{a_0^{n/2}}x\left(\frac{t}{a_0^n}\right)
\end{equation}
for all $T_1< t< T_2$, where 
$\Omega=\frac{1}{T_2-T_1}$,
\begin{equation}\label{eq:rmn}
c_{m,n}=\frac{1}{\Omega^2}e^{-jm\pi \Omega(T_1+T_2)}
\int_{-\infty}^{\infty}\left(\int_0^{\infty}\,\hat{\rho}(\omega,a)e^{j\pi\omega
(T_1+T_2)}\sinc\left(\frac{\omega}{\Omega}-m\right)\,\sinc\left(\frac{\ln a}{\ln a_0}-n
\right)\da\right)\domega,
\end{equation}
and $\hat{\rho}$ is computed in turn from $h(t,\tau)$ through (\ref{eq:h->R}).
\label{th:can}
\end{theorem}

The convergence in (\ref{eq:FSmodel}) is in the $L^2$ sense. The proof of
Theorem \ref{th:can} is included in Appendix \ref{sec:proof}.  In
(\ref{eq:FSmodel}) we see that if we receive scale-limited waveforms
over a finite time window, that we can decompose the time varying
channel into a discrete representation involving a countable sum of
weighted scale frequency shifts of the transmitted waveform. 

\section{Summary and Future Work}\label{sum}
Table~\ref{sumtable0} summarizes the projection and translation
operators used to generate the three discrete canonical channel model
discussed in this paper. Each of the models can be thought of as
sending a transmit waveform through a shaping transmission filter $Q$
and then receiving the signal through a receiving filter $P$. 
\begin{table}
\begin{center}
\[
\begin{array}{|c||c|c|c|c|}\hline
\text{model} & P & U & V & Q\\\hline
\text{time-frequency} & P_T & M_{\frac{1}{T}} & T_{\frac{1}{W}} & Q_W\\\hline
\text{time-scale} & R_{\frac{1}{2\ln a_0}} & D_{a_0} & T_{b_0} & Q_{\frac{1}{b_0}}\\\hline
\text{frequency-scale} & P_{[T_1,T_2]} & M_{1/(T_2-T_1)} & D_{a_0} & R_\frac{1}{2\ln a_0}\\\hline
\end{array}
\]
\caption{Summary of canonical models of the form $PHQ = \sum_{m,n} c_{m,n}PU^mV^nQ$.}\label{sumtable0}
\end{center}
\end{table}
The corresponding discrete channel models are presented in
Table~\ref{sumtable}. We have presented here a general theory which
generates these models based on assumptions on the transmitter and
receiver characteristics:
\begin{itemize}
\item The time-frequency model arises from:
\begin{itemize}
\item frequency bandlimited transmit waveforms
\item put through a time-frequency (narrowband) channel 
\item at a time-limited receiver.
\end{itemize}
\item The time-scale model arises from:
\begin{itemize}
\item frequency bandlimited transmit waveforms
\item put through a time-scale (wideband) channel 
\item to a scale-limited receiver.
\end{itemize}
\item The frequency-scale model arises from:
\begin{itemize}
\item scale-limited transmit waveforms
\item put through a frequency-scale channel 
\item to a time-limited receiver.
\end{itemize}
\end{itemize}

One of the many items for further study is the question of
the physical interpretation of the frequency-scale model. In what
settings can we envision a channel which imparts a limited range of
frequency and scale shifts of an input signal? Perhaps a direct path
only model of a wideband sonar signal reflecting off the undulating
surface of the ocean with a moving transmitter would impart
simultaneously a frequency shift (caused by the frequency of the ocean
surface waves) and a scale shift (caused by the change in transmission
path length during transmission). Indeed, one main topic of future
research is to characterize the channel scenarios which lead to
efficient representation in each of the three models.

\begin{table}
\begin{center}
\begin{tabular}{|c|c|}\hline
model & characterization \\\hline
time-frequency & $y(t) = \sum_{m,n} c_{m,n}e^{j2\pi mt/T}x\left(t-\frac{n}{W}\right)$ \\\hline
time-scale     & $y(t) = \sum_{m,n} c_{m,n}\frac{1}{a_0^{m/2}}x\left(\frac{t-nb_0a_0^m}{a_0^m}\right)$ \\\hline
\text{frequency-scale}& $y(t) = \sum_{m,n} c_{m,n}e^{j2\pi mt/(T_2-T_1)} \frac{1}{a_0^{n/2}}x\left(\frac{t}{a_0^n}\right)$ \\\hline

\end{tabular}
\caption{Summary of canonical models.}\label{sumtable}
\end{center}
\end{table}

Further research topics include a full analysis of the two dimensional
delay-dilation and Doppler-dilation rake receivers which arise from
these canonical models, including an analysis as to which
communication scenarios result in performance gains for the
two-dimensional rake over conventional receivers. Also, we hope to
generalize the information theoretic analysis to the delay-dilation
and Doppler-dilation rake receivers similar to that which was done for
delay-Doppler rake receiver in~\cite{sayeed99joint}. Similarly, it
would be of interest to develop a canonical time-scale and
frequency-scale multiantenna wideband channel model similar to that
proposed in~\cite{sayeed2002} for time-frequency channels.
Also,~\cite{doroslovacki96,zhang00,zhang01} introduce wavelet-based
channel models; A comparison of these models to the model derived in
this work in Section~\ref{tfrake} is a topic of future
research. Finally, we ask, is there a corresponding
underspread/overspread theory (see~\cite{matz2002LTV,matz2002TF}) for
the time-scale and frequency-scale canonical models?

\section*{Appendix}
%%%%%%%%%%%%%%%%%%%%%%%%%%%%%%%%%%%%%%%%%
%\appendix
%%%%%%%%%%%%%%%%%%%%%%%%%%%%%%%%%%%%%%%%%
 \setcounter{section}{0}%
 \setcounter{subsection}{0}%  
 \renewcommand\thesection{\Alph{section}}
%%%%%%%%%%%%%%%%%%%%%%%%%%%%%%%%%%%%%%%%%

\section{Proof of main theorem}\label{mainp}
\begin{proof}
First we expand $PQ$ using the orthonormal basis and unitary properties of the paired-up operators,
\begin{equation}
P = \sum_m\Langle\cdot,U^me_0\Rangle U^me_0
\end{equation}
and
\begin{equation}
Q = \sum_n\Langle\cdot,V^nf_0\Rangle V^nf_0,
\end{equation}
we derive,
\begin{subequations}
\begin{eqnarray}
PQx &=& \sum_m\Langle Qx,U^me_0\Rangle U^me_0\\
    &=& \sum_m\Langle \sum_n \Langle x,V^nf_0\Rangle V^nf_0,U^me_0\Rangle U^me_0\\
    &=& \sum_{m,n}\Langle x,V^nf_0\Rangle \Langle V^nf_0,U^me_0\Rangle U^me_0\label{PQdone}.
\end{eqnarray}
\end{subequations}
We use this to determine,
\begin{subequations}
\begin{eqnarray}
P\left(\sum_{m,n}c_{m,n}U^mV^n\right)Qx&=&\sum_{m,n}c_{m,n}U^mPQV^nx\label{lhs1}\\
&\hspace*{-2.2in}=&\hspace*{-1.15in}\sum_{m,n} c_{m,n} U^m\left(\sum_{k,l}\Langle V^lf_0,U^ke_0\Rangle\Langle V^nx,V^lf_0\Rangle U^ke_0\right)\label{lhs2}\\
&\hspace*{-2.2in}=&\hspace*{-1.15in}\sum_{m,n,k,l} c_{m,n}\Langle V^lf_0,U^ke_0\Rangle\Langle x,V^{-n}V^lf_0\Rangle U^mU^ke_0\label{lhs3}\\
&\hspace*{-2.2in}=&\hspace*{-1.15in}\sum_{u,s} \left(\sum_{m,n} c_{m,n}\Langle V^{n+u}f_0,U^{s-m}e_0\Rangle\right)\Langle x,V^uf_0\Rangle U^se_0\label{lhs4}
\end{eqnarray}
\end{subequations}
where the commuting property of paired-up operators was used in (\ref{lhs1}), (\ref{PQdone}) was used in moving from (\ref{lhs1}) to (\ref{lhs2}), and the unitary property of $V$ was used in moving from (\ref{lhs2}) to (\ref{lhs3}).
Now, looking to the LHS of (\ref{gthm}), we expand using the orthonormal basis and obtain,
\begin{subequations}
\begin{eqnarray}
PHQx&=&\sum_s\Langle HQx,U^se_0\Rangle U^se_0\\
&=&\sum_s\Langle H\left(\sum_u\Langle x,V^uf_0\Rangle V^uf_0\right),U^se_0\Rangle U^se_0\\
&=&\sum_{s,u}\Langle x,V^u f_0\Rangle\Langle H V^u f_0,U^s e_0 \Rangle U^s e_0\\
&=&\sum_{u,s} h_{u,s}\Langle x,V^uf_0\Rangle U^s e_0.
\end{eqnarray} 
\end{subequations}
Given $H$, we then compute,
\begin{equation}\label{hus}
h_{u,s} \defeq \Langle H V^u f_0,U^s e_0 \Rangle
\end{equation}
which we use to solve,
\begin{equation}
\sum_{m,n} c_{m,n}\Langle V^{n+u}f_0,U^{s-m}e_0\Rangle = h_{u,s},\quad\forall u,s
\end{equation}
for $c_{m,n}$. These $c_{m,n}$ satisfy  (\ref{gthm}).
\end{proof}

\section{Solving the coefficient equation}\label{ccalc}
We now discuss the form of the solution to (\ref{solvethis}). 
We define
\begin{equation}\label{akl}
a_{k,l} \defeq \Langle V^{k}f_0,U^{l}e_0\Rangle
\end{equation}
and define
\begin{equation}
\tilde{c}_{m,n} \defeq c_{n,-m}
\end{equation}
which allows us to express (\ref{solvethis}) as,
\begin{subequations}
\begin{eqnarray}
h_{u,s} &=& \sum_{m,n} c_{m,n}\Langle V^{n+u}f_0,U^{s-m}e_0\Rangle \\
        &=& \sum_{m,n} \Langle V^{u-n}f_0,U^{s-m}e_0\Rangle \tilde{c}_{n,m} \\
        &=& \left(a\star\tilde{c}\right)_{u,s}\label{ceq}
\end{eqnarray}
\end{subequations}
where 
\begin{equation}
\left(a\star\tilde{c}\right)_{u,s} \defeq \sum_{k,l}a_{u-k,s-l}\tilde{c}_{k,l} = \sum_{k,l}a_{k,l}\tilde{c}_{u-k,s-l}
\end{equation}
Expressing $h$, $a$, and $\tilde{c}$ in the Z-transform domain,
\begin{eqnarray}
A(z_1,z_2)         & \defeq \sum_{k,l} z^k_1z^l_2a_{k,l} & = \sum_{k,l} z^k_1z^l_2\Langle V^{k}f_0,U^{l}e_0\Rangle\label{capA}\\
H(z_1,z_2)         & \defeq \sum_{k,l} z^k_1z^l_2h_{k,l} & = \sum_{k,l} z^k_1z^l_2\Langle H V^k f_0,U^l e_0 \Rangle\\
\tilde{C}(z_1,z_2) & \defeq \sum_{k,l} z^k_1z^l_2\tilde{c}_{k,l}&
\end{eqnarray}
we can write (\ref{ceq}) as,
\begin{equation}
H=A\tilde{C}
\end{equation}
and solve for $\tilde{C}$
\begin{equation}
\tilde{C}(z_1,z_2) = \frac{H(z_1,z_2)}{A(z_1,z_2)}.
\end{equation}
In terms of $c_{m,n}$, this is,
\begin{equation}
c_{m,n} = Z^{-1}\left(\frac{H(z_1,z_2)}{A(z_1,z_2)}\right)_{-n,m}\label{C}
\end{equation}
where 
\begin{equation}
Z^{-1}\left(F(z_1,z_2)\right)_{m,n}=\int^1_0\int^1_0 e^{-j2\pi\theta_1 m}e^{-j2\pi\theta_2 n}F\left(e^{j2\pi\theta_1},e^{j2\pi\theta_2}\right)\dtheta_1\dtheta_2 
\end{equation}
We can express (\ref{C}) as a convolution of coefficients by defining
\begin{equation}
\hat{A}(e^{j2\pi\theta_1},e^{j2\pi\theta_2}) \defeq \frac{1}{A(e^{j2\pi\theta_1},e^{j2\pi\theta_2})}\label{Ahat}
\end{equation}
and
\begin{equation}
\hat{a}_{m,n} \defeq \int^1_0\int^1_0 e^{-j2\pi\theta_1 m}e^{-j2\pi\theta_2 n}\hat{A}\left(e^{j2\pi\theta_1},e^{j2\pi\theta_2}\right)\dtheta_1\dtheta_2 ,\label{ahatd}
\end{equation}
and we can obtain the $c_{m,n}$ using
\begin{equation}
c_{m,n}=\tilde{c}_{-n,m} = (\hat{a}\star h)_{-n,m}.\label{thisone}
\end{equation}

\subsection*{Coefficient calculation}
Thus, to calculate the coefficients $c_{m,n}$,
\begin{enumerate}
\item calculate $h_{k,l}$ via (\ref{hus}),
\item calculate $a_{m,n}$ via (\ref{akl}),
\item use $a_{m,n}$ to obtain $A(e^{j2\pi\theta_1},e^{j2\pi\theta_2})$ via (\ref{capA}),
\item use $A(e^{j2\pi\theta_1},e^{j2\pi\theta_2})$ to obtain $\hat{a}_{m,n}$ vua (\ref{Ahat}) and (\ref{ahatd}), and
\item use $h_{k,l}$ and $\hat{a}_{m,n}$ to obtain $c_{m,n}$ via (\ref{thisone}).
\end{enumerate}

Here, we present the highlights of the coefficient calculation procedure for the time-frequency and time-scale canonical models.  For more detailed steps, consult~\cite{rickardphd}.
\subsection*{Example: time-frequency}
\begin{equation}
h_{k,l} = \sqrt{\frac{W}{T}}\tripint 
1_{[0,T]}(t)e^{j2\pi
t(\theta-\frac{l}{T})}\sinc(Wt-k-W\tau)S(\theta,\tau)\dtheta\dtau\dt
\end{equation}
\begin{equation}
a_{m,n} = \sqrt{\frac{W}{T}}\int^T_0 e^{-j2\pi\frac{nt}{T}}\sinc(Wt-m)\dt
\end{equation}
For $\theta_1,\theta_2\in[0,1]$,
\begin{equation}
A(e^{j2\pi\theta_1},e^{j2\pi\theta_2})=\left\{
\begin{array}{r@{\quad:\quad}l} \sqrt{WT}e^{j2\pi WT\theta_1\theta_2}&
\theta_1\in\left(0,\frac{1}{2}\right) \\ \sqrt{WT}e^{j2\pi
WT(\theta_1-1)\theta_2} &
\theta_1\in\left(\frac{1}{2},1\right)\end{array}\right.
\end{equation}
\begin{equation}
\hat{a}_{m,n}=\frac{1}{\sqrt{WT}}\int^1_0 e^{-j2\pi\theta_2 n}\sinc(WT\theta_2+m)=\frac{1}{WT}a_{-m,n}\dtheta_2
\end{equation}
\begin{equation}
c_{m,n} = \doubint S(\theta,\tau) e^{j\pi(T\theta+m)}\sinc(T\theta+m) \sinc(n+W\tau)  \dtheta\dtau 
\end{equation}
which are precisely the coefficients in (\ref{cantfrake}).

\subsection*{Example: time-scale}
\begin{equation}
h_{u,s}=\frac{1}{\sqrt{b_0\ln a_0}}\doubint\frac{1}{\sqrt{|a|}}\Lop(a,b)\left(\int^\infty_0\frac{1}{\sqrt{t}}\sinc\left(\frac{t-b}{ab_0}-u\right)\sinc\left(\frac{\ln t}{\ln a_0}-s\right)\dt\right)\da\db
\end{equation}
\begin{equation}
a_{m,n} = \sqrt{\frac{1}{b_0\ln a_0}}\int^\infty_0 \frac{1}{\sqrt{t}}\sinc\left(\frac{t}{b_0}-m\right)\sinc\left(\frac{\ln|t|}{\ln a_0}-n\right)\dt
\end{equation}
For $\theta_1,\theta_2\in\left[-\frac{1}{2},\frac{1}{2}\right]$, in distributional sense,
\begin{equation}
A(\theta_1,\theta_2)= \sqrt{\frac{1}{b_0\ln a_0}}b_0^{\frac{1}{2}+j2\pi\frac{\theta_2}{\ln a_0}}\int^\infty_0 t^{-\frac{1}{2}+j2\pi\frac{\theta_2}{\ln a_0}}e^{j2\pi\theta_1t}\dt
\end{equation}
\begin{equation}
\hat{a}_{m,n} =\sqrt{\ln a_0} \int^{\frac{1}{2}}_{-\frac{1}{2}}\int^{\frac{1}{2}}_{-\frac{1}{2}}\frac{b_0^{-j2\pi\frac{\theta_2}{\ln a_0}}e^{-j2\pi\theta_1 m}e^{-j2\pi\theta_2 n}}{\int^\infty_0 t^{-\frac{1}{2}+j2\pi\frac{\theta_2}{\ln a_0}}e^{j2\pi\theta_1t}\dt}\dtheta_1\dtheta_2
\end{equation}
\begin{equation}
c_{m,n} = \doubint\Lop(a,b)\sinc\left(m-\frac{\ln a}{\ln a_0}\right)\sinc\left(n-\frac{b}{ab_0}\right)\da\db.
\end{equation}

\section{The equivalence between (\ref{eq:1}) and (\ref{eq:8})}\label{sec:equiv}

In this section we obtain the correspondence relations between the two
forms (\ref{eq:1}) and (\ref{eq:8}) of a general time-varying linear system
when input signal are supported on positive time domain, and the
observation is restricted to a positive time horizon.

Consider first the input-output relationship given by (\ref{eq:1}). For
positive time supported input signals, the output is given by
\[ y(t) = \int_{0}^{\infty} h(t,t-\tau)x(\tau)\dtau \]
We change the integration variable $\tau\rightarrow \frac{t}{a}$, and since
we have a positive time horizon, i.e. $t>0$, we obtain:
\[ y(t) = \int_0^{\infty}h\left(t,t-\frac{t}{a}\right)x\left(\frac{t}{a}\right)\frac{t}{a^2}\,d\,a \]
Now denote $\rho(t,a)=\frac{t}{a\sqrt{a}}h\left(t,t-\frac{t}{a}\right)$, and $\hat{\rho}(\omega,a)$
its Fourier transform with respect to $t$. Then the inverse Fourier transform
allows us to write
\[ y(t) = \int_{-\infty}^{\infty}\left(\int_0^{\infty}\hat{\rho}(\omega,a)e^{j2\pi \omega t}
\frac{1}{\sqrt{a}}x\left(\frac{t}{a}\right)\da\right)\domega \]
that is (\ref{eq:8}), where, explicitly,
\begin{equation}
\label{eq:h->R}
\hat{\rho}(\omega,a) = \frac{1}{a\sqrt{a}}
\int_{0}^{\infty} e^{-j2\pi \omega t}t\,h\left(t,t\frac{a-1}{a}\right)\,\dt 
\end{equation}

For the converse, assume the input-output relationship in given by
(\ref{eq:8}). Then, performing the integration over $\omega$ first
we obtain
\[ y(t) = \int_0^{\infty} \rho(t,a)\frac{1}{\sqrt{a}}x\left(\frac{t}{a}\right)\da.\]
Next we need to change the integration variable $a$ into $\tau=t-\frac{t}{a}$
\[ y(t) = \int_{-\infty}^{t} \rho\left(t,\frac{t}{t-\tau}\right)\sqrt{\frac{t-\tau}{t}}
 x(t-\tau) \dtau \]
which is exactly (\ref{eq:1}) with
\begin{equation}
\label{eq:R->h}
h(t,\tau) = 1_{t>\tau}(t)\sqrt{\frac{t-\tau}{t}}\int_{-\infty}^{\infty}e^{j2\pi 
\omega t}\hat{\rho}\left(\omega,\frac{t}{t-\omega}\right)\,\domega
\end{equation}
where $1_{t>\tau}(t)\defeq 1_{[\tau,\infty)}(t)$.
\section{Proof of Theorem \ref{th:can}}\label{sec:proof}
 We follow the recipe proposed in Appendix~\ref{ccalc}. The two sets of paired-up operators and generators are: $(P=1_{[T_1,T_2]},M_{\Omega},e_0(t)=\sqrt{\Omega}
1_{[T_1,T_2]}(t))$ and $(Q=\M^{-1}1_{[-\frac{1}{2\ln a_0},\frac{1}{2\ln a_0}]}\M,
D_{a_0}, f_0(t)=\sqrt{\frac{1}{t\ln a_0}}\sinc(\frac{\ln t}{\ln a_0})1_{t>0}(t))$.
First we need to compute $h_{k,l}$ and $a_{m,n}$. We have:
\[ h_{k,l}=\ip{HD_{a_0}^kf_0}{M_{\Omega}^le_0} = \frac{1}{\sqrt{\ln a_0}}
\int_{-\infty}^{\infty}\left(\int_0^{\infty}\hat{\rho}(\omega,a)\left(\int_{T_1}^{T_2}
\frac{1}{\sqrt{t}}e^{j2\pi t(\omega-l\Omega)}\sinc\left(\frac{\ln t/a}{\ln a_0}-k\right)\dt\right)\da\right)\domega \]
\[ a_{m,n} = \ip{D_{a_0}^mf_0}{M_{\Omega}^ne_0} = \frac{1}{\sqrt{\ln a_0}}\int_{T_1}^{T_2}
\frac{1}{\sqrt{t}}e^{-j2\pi \Omega nt}\sinc\left(\frac{\ln t}{\ln a_0}-m\right)\dt \]
Next we compute $A(z_1,z_2)=\sum_{m,n}a_{m,n}z_1^mz_2^n$,
 at $z_1=e^{j2\pi \theta_1}$, $z_2=e^{j2\pi \theta_2}$ for 
$\theta_1\in[-\frac{1}{2},\frac{1}{2}]$, $\theta_2\in[0,1]$. We obtain:
\[ A(e^{j2\pi \theta_1},e^{j2\pi \theta_2})=\sqrt{\frac{\Omega}{(\theta_2+n_0)\ln a_0}}e^{j2\pi \theta_1\frac{1}{\ln a_0}\,\ln\frac{\theta_2+n_0}{\Omega}} \]
where $n_0=n_0(\theta_2)$ is the only integer so that 
$\frac{\theta_2+n_0}{\Omega}\in[T_1,T_2)$. Then
\[ \hat{A}(e^{j2\pi \theta_1},e^{j2\pi \theta_2})=
\frac{1}{A(e^{j2\pi \theta_1},e^{j2\pi \theta_2})}=
\sqrt{\frac{(\theta_2+n_0)\ln a_0}{\Omega}}
e^{-j2\pi \theta_1\frac{1}{\ln a_0}\,\ln\frac{\theta_2+n_0(\theta_2)}{\Omega}} \]
which has its Fourier expansion with coefficients $\hat{a}_{m,n}$ given by
\[ \hat{a}_{m,n}=\sqrt{\ln a_0}\int_{T_1}^{T_2}\sqrt{t}
e^{-j2\pi  n\Omega t}\sinc\left(m+\frac{\ln t}{\ln a_0}\right)\dt \]
Then the coefficients $c_{m,n}$ that solve the equation $a\star\tilde{r}=h$
with $\tilde{r}_{n,m}=c_{m,-n}$ are given by
\[ c_{m,n}=(\hat{a}\star h)_{-n,m}=\frac{1}{\Omega^2}e^{-jm\pi \Omega(T_1+T_2)}
\int_{-\infty}^{\infty}\left(\int_0^{\infty}\hat{\rho}(\omega,a)e^{j\pi\omega
(T_1+T_2)}\sinc\left(\frac{\omega}{\Omega}-m\right)\,\sinc\left(\frac{\ln a}{\ln a_0}-n
\right)\da\right)\domega \]
which is exactly (\ref{eq:rmn}).

\bibliographystyle{unsrt} 
\bibliography{phdthesis}

\end{document}